# Graph neural networks and attention-based CNN-LSTM for protein classification

Zhuangwei Shi[1,2,3,*] and Bo Li[2,4]

1. College of Artificial Intelligence, Nankai University, Tianjin, China. Email: zwshi@mail.nankai.edu.cn
2. Shenzhen Research Institute, Nankai University, Shenzhen, Guangdong, China
3. Ji Hua Laboratory, Foshan, Guangdong, China. Email: shizw@jihualab.ac.cn
4. State Key Laboratory of Medicinal Chemical Biology, College of Pharmacy, Nankai University, Tianjin, China

## Abstract

This paper focuses on three critical problems on protein classification. Firstly, Carbohydrate-active enzyme (CAZyme) classification can help people to understand the properties of enzymes. However, one CAZyme may belong to several classes. This leads to Multi-label CAZyme classification. Secondly, to capture information from the secondary structure of protein, protein classification is modeled as graph classification problem. Thirdly, compound-protein interactions prediction employs graph learning for compound with sequential embedding for protein. This can be seen as classification task for compound-protein pairs. This paper proposes three models for protein classification. Firstly, this paper proposes a Multi-label CAZyme classification model using CNN-LSTM with Attention mechanism. Secondly, this paper proposes a variational graph autoencoder based subspace learning model for protein graph classification. Thirdly, this paper proposes graph isomorphism networks (GIN) and Attention-based CNN-LSTM for compound-protein interactions prediction, as well as comparing GIN with graph convolution networks (GCN) and graph attention networks (GAT) in this task. The proposed models are effective for protein classification. Source code and data are available at https://github.com/zshicode/GNN-AttCL-protein. Besides, this repository collects and collates the benchmark datasets with respect to above problems, including CAZyme classification, enzyme protein graph classification, compound-protein interactions prediction, drug-target affinities prediction and drug-drug interactions prediction. Hence, the usage for evaluation by benchmark datasets can be more conveniently.

## Keywords

Variational graph autoencoder, CNN-LSTM, Attention mechanism, enzyme protein classification

# Background

Classification of enzyme proteins, is important for enzyme research. Carbohydrate-active enzymes (CAZymes) are important to bioenergy, human gut microbiome, plant pathogen and so on. Carbohydrate-active enzyme (CAZyme) classification can help people to understand the properties and functions of enzymes.

CAZymes belong to six classes/superfamilies: glycosyltransferases (GTs), glycoside hydrolases (GHs), polysaccharide lyases (PLs), carbohydrate esterases (CEs), enzymes of auxiliary activities (AAs) and carbohydrate-binding modules (CBMs). (Xu et al. 2020)

However, one CAZyme may belong to several classes. For instance, CAZyme AGF60570.1 belongs to CBM13 and GH59. This leads to Multi-label Carbohydrate-active enzyme (CAZyme) classification. Here, label vector is defined as (CE, GT, GH, PL, AA, CBM). For instance, the label vector of AGF60570.1 can be defined as (0,0,1,0,0,1).

Borgwardt et al. (2005) designed protein graph models to contain information about structure, sequence and chemical properties of a protein. For this purpose, each graph represents exactly one protein. Nodes in graph represent secondary structure elements (SSEs) within the protein structure, i.e. helices, sheets and turns. 85% of secondary structure elements are in these three categories: $\alpha$-helix, $\beta$-sheet and $\beta$-turn. In addition to the above regular conformations, some irregular conformations of the peptide plane in the polypeptide chain are called random coil. Node feature vector is the one-hot encoding of these three categories SSEs: $\alpha$-helix, $\beta$-sheet and $\beta$-turn. Edges connect nodes if those are neighbors along the AA sequence or if they are neighbors in space within the protein structure.

When two nodes are less than a threshold measured by angstroms apart, connect them with an edge. Borgwardt et al. (2005) proposed ENZYMES dataset and PROTEINS dataset for protein classification.

DD dataset (Dobson et al. 2003) is a dataset containing 1178 proteins. To designed protein graph models to contain information about structure, sequence and chemical properties of a protein, each graph represents exactly one protein. Nodes in graph represent amino acids. Node embedding contains information about residue, sequence and chemical properties of an amino acid, and the dimension is 89. When two nodes are less than a threshold measured by angstroms apart 6 angstroms apart (i.e. 6e-10 meters), connect them with an edge.

Biological sequence and graph classification tasks are widely applied in bioinformatics research. Compounds, RNAs and proteins can all be modeled as graphs for graph learning. Graph neural networks are widely adopted for modeling molecular graphs (Yang et al. 2019; Xu et al. 2021). XGBoost is used for modeling drug–target binding affinities (He et al. 2017). Graph neural networks with XGBoost classifier was used for molecular property prediction (Deng et al. 2021).

Compound-protein interactions (also known as drug-target affinities, drug-protein interactions, ligand-protein interactions etc.) prediction is important for biochemical research. The task integrates compound graphs (for graph learning) and protein sequences (for sequence learning) to conduct machine learning techniques for prediction (Tsubaki et al. 2019; Quan et al. 2019; Chen et al. 2020). RNA classification can be modeled as both sequence classification and graph classification (Rossi et al. 2019).

Note that some datasets model compound-protein interaction (drug-target affinity) prediction as regression task. However, the drug-target affinity prediction also integrates drug compound graphs (for graph learning especially using GNN) and protein sequences (for sequence learning) to conduct machine learning techniques for prediction (Nguyen et al. 2021; Huang et al. 2021; Yuan et al. 2022; Yang et al. 2022).

Drug-target affinities and drug-drug interactions research on the drug for COVID-19 therapy is essential. DRKG (https://github.com/gnn4dr/DRKG/) is Drug Repurposing Knowledge Graph for COVID-19. Chen et al. (2021) adopted DRKG dataset for Drug-drug interactions prediction.

This paper focuses on critical problems on protein classification:

1. This paper proposes a Multi-label CAZyme classification model using CNN-LSTM with Attention mechanism.
2. This paper proposes a variational graph autoencoder based subspace learning model for protein graph classification.
3. This paper proposes Graph Isomorphism Networks and Attention-based CNN-LSTM for compound-protein interactions prediction. Ablation study is proposed to validate that graph isomorphism networks is superior to graph convolution and graph attention while incorporating Attention-based CNN-LSTM for compound-protein interactions prediction.
4. Source code and data are available at https://github.com/zshicode/GNN-AttCL-protein. Evaluation shows that the proposed models are effective for protein classification.
5. This GitHub repository collects and collates the benchmark datasets for three problems mentioned above, including CAZyme classification, enzyme protein graph classification, compound-protein interactions prediction, drug-target affinities prediction and drug-drug interactions prediction. Hence, the usage for evaluation by benchmark datasets can be more conveniently.

# Materials and Methods

## CLACAZy: Multi-label CAZyme classification model using CNN-LSTM with Attention mechanism

CNN can capture salient (local) features and LSTM can capture sequential (global) features. Incorporating attention mechanism to multi-scale CNN (Jin et al. 2021), can capture both local and global features of protein sequences efficiently.

Inspired by this, we propose a Multi-label Carbohydrate-active enzyme (CAZyme) classification model using CNN-LSTM with Attention mechanism. The model first adopts Word2Vec for extracting protein embeddings. This tool is also known as BioVec (https://pypi.org/project/biovec/). (Asgari et al. 2015) This can be considered as pretraining process, which is also widely used for protein classification (Jin et al. 2022).

Then, the embeddings are sequentially put into multi-scale convolution layer, BiLSTM layer and self-attention layer. Finally, after being put into a fully connected layer with sigmoid activation function, the model obtains the classification results. This model is so called CLACAZy.

The dbCAN2 (https://bcb.unl.edu/dbCAN2/index.php) database (Zhang et al. 2018) collected 1,066,327 CAZymes until July 31, 2018. The data can be downloaded via the website of dbCAN2 (https://bcb.unl.edu/dbCAN2/download/CAZyDB.07312018.fa).

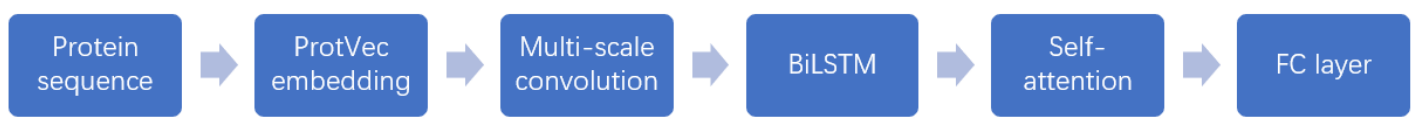

Fig. 1 CLACAZy

# VSPool: Variational graph autoencoder based subspace learning model for enzyme protein classification

Vanilla GNN focus on classification task upon nodes on a graph. The graph convolutional networks (GCN) (Kipf et al. 2017), graph autoencoders (GAE) (Kipf et al. 2016), variational graph autoencoders (VGAE) (Kipf et al. 2016), graph attention networks (GAT) (Velikovic et al. 2018), graph isomorphism network (GIN) (Xu et al. 2019) and many other models were proposed. Probabilistic theory inspired graph neural networks (Shi et al. 2021), such as varitional inference and conditional random field, are powerful for detecting latent variable dependency and learning efficient representation.

Graph neural networks do not only adopt on node classification, but also adopt on graph classification or graph pooling. This task focuses on representation learning from the whole graph. Basic methods were based on graph kernel (Shervashidze et al. 2011). Novel GNN for node classification, such as GCN and GIN, can adopt for graph classification by adding mean pooling operator that outputs the mean vector of all feature vectors of nodes. However, researchers have proposed a series of graph neural network models mainly for graph pooling and graph classification tasks, such as Diffpool (Ying et al. 2018) and graph u-net (Gao et al. 2019).

This paper proposes a VGAE based subspace learning model for enzyme protein classification, named VSPool. It uses a variational graph autoencoder to map the features of nodes $X \in \mathbb{R}^{n \times m}$ to subspace vector $v \in \mathbb{R}^{n \times 1}$, and then adopt $v$ to reconstruct features $X' \in \mathbb{R}^{n \times m}$. Note that

$$u = X^T v \in \mathbb{R}^{m \times 1}, \forall n \tag{1}$$

Output $u$ can be the representation of a graph. The loss function is the sum of reconstruction error, classification error and Kullback-Leibler divergence.

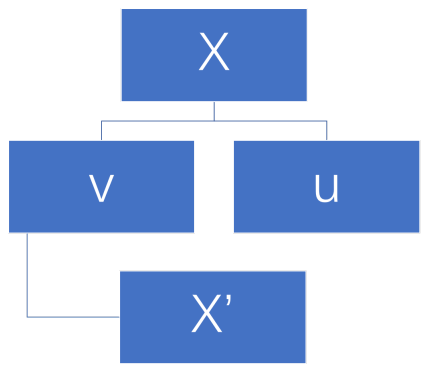

Fig. 2 VSPool

Table 1 lists the situation on protein graph classification datasets, where AvgNodes and AvgEdges denote the average NodeNumber and EdgeNumber of graphs in the datasets. GNum, CNum, PNum, NNum denote number of graphs, classes, positive samples and negative samples, respectively. NFDim denotes node feature dimension.

ENZYMES is a 6-class classification dataset with 600 graphs, each class contains 100 graphs. The protein classification datasets, as well as other graph classification datasets towards compounds etc., can be downloaded from (https://ls11-www.cs.tu-dortmund.de/people/morris/graphkerneldatasets/) or (https://chrsmrrs.github.io/datasets/docs/datasets/).

Table 1 Situation on protein graph classification datasets

| Name | GNum | CNum | PNum | NNum | NFDim | AvgNodes | AvgEdges |
|------|------|------|------|------|-------|----------|----------|
| DD | 1178 | 2 | 691 | 487 | 89 | 284.32 | 715.66 |
| PROTEINS | 1113 | 2 | 663 | 450 | 3 | 39.06 | 72.82 |
| ENZYMES | 600 | 6 | - | - | 3 | 32.63 | 62.14 |

# GACLCPI: Graph Isomorphism Networks and Attention-based CNN-LSTM for compound-protein interactions prediction

## Compound-protein interactions prediction

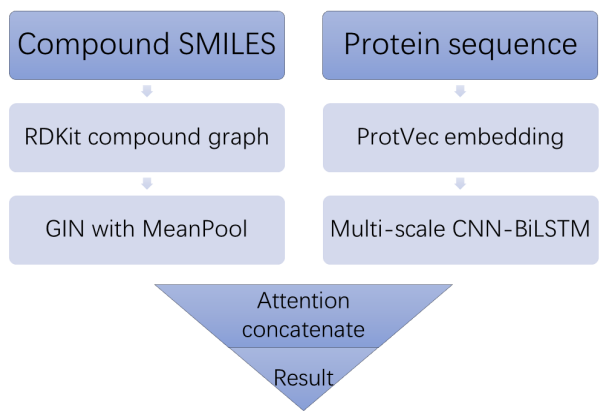

Fig 3. GACLCPI

Algorithm 1 GACLCPI

Require: Compound SMILES, protein sequence

Ensure: Prediction of compound-protein pair

1. Use RDKit (http://www.rdkit.org/) to extract compound graph. Nodes denote the atoms of a compound. Edges denote chemical bonds. Node features are one-hot encoding of atom elements (C, N, H, O, ...)
2. Use ProtVec for protein sequence embedding (Asgari et al. 2015).
3. Use Graph Isomorphism Networks with MeanPooling (Xu et al. 2019) for compound feature vector learning.
4. Use multi-scale CNN-LSTM (Su et al. 2019) for protein feature vector learning.
5. Use attention mechanism for concatenating compound and protein feature vectors (Tsubaki et al. 2019).

Return: Concatenating vector as result

Following prior work (Tsubaki et al. 2019), we used compound-protein interaction dataset called human for evaluation. The human dataset contains 3369 interaction-pairs and 3369 non-interaction pairs between 1052 compounds and 852 human proteins.

The human dataset and C.elegans dataset from (Tsubaki et al. 2019; Quan et al. 2019) consider compound-protein interaction prediction as binary classification task. The two datasets contains subdatasets with different positive:negative samples ratios (1:1,1:3,1:5). For example, the 1:1 human dataset contains 3369 interaction-pairs and 3369 non-interaction pairs between 1052 compounds and 852 human proteins. The 1:3 human dataset contains 3369 interaction-pairs and 10107 non-interaction pairs between 1052 compounds and 852 human proteins.

Ablation study is proposed to validate that graph isomorphism networks is superior to graph convolution and graph attention while incorporating Attention-based CNN-LSTM for compound-protein interactions prediction, see Results.

# Drug-target affinities prediction

The DAVIS and KIBA datasets from (He et al. 2017; Ozturk et al. 2018; Yuan et al. 2022) consider compound-protein interaction (drug-target affinity) prediction as regression task. Specially, Huang et al. (2021) modeled DAVIS and KIBA datasets for inferring drug-target affinities as classification task.

DAVIS dataset contains interaction between 68 compounds and 442 proteins. KIBA dataset contains interaction between 2111 compounds and 229 proteins.

The machine learning models aims to fit some chemical metrics of drug-target pairs, e.g. dissociation constant $K_d$ or its negative logarithmic form (in DAVIS dataset) (see He et al. 2017)

$$pK_d = -\log_{10}(K_d/10^9) \qquad (2)$$

Both of classification task and regression task modeling this problem, can be seen as representation learning for compound-protein (drug-target, ligand-protein) pairs. So these tasks in different applied perspective can be unified into same framework (Yang et al. 2022).

The above framework denotes a compound-protein pair as a sample. However, affinity values can also be storaged as association matrix. The matrix shape among $m$ compounds and $n$ proteins, is $F \in \mathbb{R}^{m \times n}$. This step can formulate the drug-target affinity regression task as matrix factorization for biological association prediction (Liu et al. 2020; Shi et al. 2021).

## Drug-drug interactions prediction

Drug-drug interactions prediction framework denotes drug-drug pair as a sample, to employ binary classification (Huang et al. 2020; Chen et al. 2021; Wang et al. 2021) or multi-class classification (Chen et al. 2021). Each drug compound is a graph. However, label values can also be storaged as association matrix. The matrix shape among $n$ drugs, is $A \in \mathbb{R}^{n \times n}$. This step can formulate the Drug-drug interactions prediction task as matrix factorization for biological association prediction.

Wang et al. (2021) adopted three datasets for Drug-drug interactions (DDI) binary classification prediction.

- ZhangDDI dataset contains 548 drugs and 48,548 pairwise DDIs. Wang et al. (2021) removed the data items that cannot be converted into graphs from SMILES strings. And it remains 544 drugs and 40,255 interactions.
- ChCh-Miner dataset, also named as BIOSNAP dataset in Huang et al. (2020), comes from Stanford Biomedical Network Dataset Collection (BIOSNAP) (http://snap.stanford.edu/biodata/index.html). It contains 1,514 drugs and 48,514 DDI links. Wang et al. (2021) removed the data items that cannot be converted into graphs from SMILES strings. And it remains 997 drugs.
- DeepDDI contains 192,284 DDIs from DrugBank (https://go.drugbank.com/). It calls DrugBank dataset in (Huang et al. 2020; Chen et al. 2021). Wang et al. (2021) removed the data items that cannot be converted into graphs from SMILES strings. And it remains 1,704 drugs.

Chen et al. (2021) adopted DRKG dataset (https://github.com/gnn4dr/DRKG/) for Drug-drug interactions (DDI) binary classification and multi-class classification prediction. DRKG is Drug Repurposing Knowledge Graph for COVID-19. In binary classification DDI dataset, it contains 2,322 drugs and 1,178,210 pairwise DDIs. In multi-class classification DDI dataset, it contains 2,322 drugs and 172,406 pairwise DDIs.

# Usage

## CLACAZy

The dbCAN2 (https://bcb.unl.edu/dbCAN2/index.php) database (Zhang et al. 2018) collected 1,066,327 CAZymes until July 31, 2018. The data can be downloaded via the website of dbCAN2 (https://bcb.unl.edu/dbCAN2/download/CAZyDB.07312018.fa).

We select 4683 CAZymes from all 1,066,327 CAZymes, to construct the dataset. 3759 CAZymes are divided into training set, while the other 924 CAZymes are divided into testing set. See `train.fasta` and `test.fasta` .

The code can be run through

```
cd CLACAZy
python main.py
```

Hyper parameter setting can refer to our code repository.

## VSPool

Data formed for VSPool experiments can be downloaded from this link: (https://github.com/HongyangGao/Graph-U-Nets/tree/master/data) (Gao et al. 2019).

After downloading the data, the code can be run through

```
cd VSPool
```

Type

```
./run_GNN.sh DATA FOLD GPU
```

to run on dataset using fold number (1-10).

You can run

```
./run_GNN.sh DD 0 0
```

to run on DD dataset with 10-fold cross validation on GPU #0.

# GACLCPI

## Compound-protein interactions prediction

`GACLCPI/dataset/human` includes the human dataset from
https://github.com/masashitsubaki/CPI_prediction (Tsubaki et al. 2019) or
https://github.com/XuanLin1991/GraphCPI. (Quan et al. 2019) This dataset models compound-
protein interaction prediction as binary classification task.

The code (on human dataset for classification task) can be run through

```
cd GACLCPI/code
python preprocess_data.py
python run_training.py
```

The code implements GCN, GAT and GIN for compound embedding (default is GIN), in function `gnn`
of class `CompoundProteinInteractionPrediction`.

```
class CompoundProteinInteractionPrediction(nn.Module):
    def __init__(self):
        super(CompoundProteinInteractionPrediction, self).__init__()
        # ...

    def gnn(self, xs, A, layer, model=GNNMODEL):
        if model=='GCN':
            # graph convolution
        elif model=='GAT':
            # graph attention
        else:
            # default is GIN
```

## Drug-target affinities prediction

`GACLCPI/data/davis` and `GACLCPI/data/kiba` include the DAVIS and KIBA datasets from
https://github.com/hetong007/SimBoost (He et al. 2017) (by R language) or
https://github.com/hkmztrk/DeepDTA/tree/master/data (Ozturk et al. 2018) or
https://github.com/yuanweining/FusionDTA. (Yuan et al. 2022) These datasets model compound-
protein interaction (drug-target affinity) prediction as regression task.

Both of classification task and regression task modeling this problem, can be seen as representation
learning for compound-protein (drug-target, ligand-protein) pairs. Each pair denotes a sample. So
these tasks in different applied perspective can be unified into same framework.

The script `GACLCPI/data.py` preprocesses the DAVIS and KIBA datasets for regression task.

- The DAVIS and KIBA datasets are transformed into the list of compound-protein pairs (in CSV files).
- The lists of compounds (including name and SMILE of each compound) and proteins (including name and sequence of each protein) are created (in CSV files). Protein names and sequences are also formatted as FASTA files.
- The affinity values are storaged as matrices. The matrix shape is `CompoundNum*ProteinNum`. This step can formulate the drug-target affinity regression task as matrix factorization for biological association prediction.

## Drug-drug interactions prediction

`GACLCPI/ddi/miracle_ddi.py` transforms datasets in MIRACLE model (Wang et al. 2021) (https://github.com/isjakewong/MIRACLE) for DDI. The data folder is (https://github.com/isjakewong/MIRACLE/tree/main/MIRACLE/datachem). The labels are transformed to be storaged as matrices. The matrix shape is `DrugNum*DrugNum`. This step can formulate the Drug-drug interactions prediction task from drug-drug pair classification (such that a pair denotes a sample) to matrix factorization.

Chen et al. (2021) adopted MUFFIN model (https://github.com/xzenglab/MUFFIN) that uses DRKG dataset (https://github.com/gnn4dr/DRKG/) for Drug-drug interactions (DDI) binary classification and multi-class classification prediction. DRKG is Drug Repurposing Knowledge Graph for COVID-19.

`GACLCPI/ddi/drkg_ddi.py` transforms DRKG dataset for DDI. The data folder is (https://github.com/xzenglab/MUFFIN/tree/main/data/DRKG). The labels are transformed to be storaged as matrices. The matrix shape is `DrugNum*DrugNum`.

# Requirements

The code has been tested running under Python 3.7.4, with the following packages and their dependencies installed:

```
numpy
pytorch
sklearn
biopython
biovec
rdkit
```

# Results

## CLACAZy

The prediction results are listed as below. The results and case study show that the proposed CLACAZy outperforms baseline methods.

Table 2 AUROC and AUPR (mean+-std)

| Model | AUROC | AUPR |
|---|---|---|
| Logistic Regression | 0.7123 +- 0.0102 | 0.2337 +- 0.0049 |
| Random Forest | 0.7438 +- 0.0212 | 0.3081 +- 0.0182 |
| SVM | 0.7538 +- 0.0194 | 0.3476 +- 0.0243 |
| Multi-layer Perceptron | 0.7805 +- 0.0332 | 0.3969 +- 0.0472 |
| CLACAZy | 0.8151 +- 0.0327 | 0.4573 +- 0.0546 |

Table 3 Case study. CAZyme AGF60570.1 belongs to CBM13 and GH59. LR, RF and SVM achieve accuracy 3/6. MLP achieves accuracy 5/6. CLACAZy achieves accuracy 6/6.

| Model | CE | GT | GH | PL | AA | CBM | MSE |
|---|---|---|---|---|---|---|---|
| LR | 0.5502 | 0.2489 | 0.2178 | 0.0972 | 0.1128 | 0.2387 | 0.2630 |
| RF | 0.5732 | 0.3195 | 0.3389 | 0.3560 | 0.1387 | 0.4038 | 0.2281 |
| SVM | 0.5346 | 0.3703 | 0.2738 | 0.3429 | 0.2157 | 0.4452 | 0.2370 |
| MLP | 0.1832 | 0.4463 | 0.5752 | 0.6239 | 0.1932 | 0.5297 | 0.1768 |
| CLACAZy | 0.1176 | 0.2474 | 0.6783 | 0.2817 | 0.1092 | 0.5932 | 0.0725 |

# VSPool

Predicting whether these proteins are enzymes in DD datasetis a binary classification problem. VSPool is compared with GCN with mean pooling, GIN with mean pooling, Diffpool, and graph u-net.

Table 4 VSPool outperforms all compared methods.

| Method | Acc |
|---|---|
| WL Graph Kernel | 0.7834 |
| GCN | 0.7590 |
| GIN | 0.7530 |
| DiffPool | 0.8064 |
| Graph U-Nets | 0.8243 |
| VSPool | 0.8275 |

# GACLCPI

Table 5 AUROC and AUPR

| Model | AUROC | AUPR |
|---|---|---|
| kNN | 0.860 | 0.931 |
| Logistic Regression | 0.940 | 0.902 |
| Random Forest | 0.911 | 0.919 |
| SVM | 0.910 | 0.972 |
| DeepCPI (Tsubaki et al. 2019) | 0.970 | 0.947 |
| GACLCPI | 0.975 | 0.970 |

Table 6 Ablation study

| Model | AUROC | AUPR |
|---|---|---|
| GACLCPI-GCN | 0.969 | 0.935 |
| GACLCPI-GAT | 0.972 | 0.957 |
| GACLCPI-GIN | 0.975 | 0.970 |

Following prior works (Tsubaki et al. 2019; Yuan et al. 2022), case study was adopted for illustrating the biochemical meaning of the proposed model. Fig. 4 shows examples of interactions between a drug compound and a protein. The 3D visuallization is obtained via Protein Data Bank in Europe (https://www.ebi.ac.uk/pdbe). Complex is highlighted with the detection by the attention mechanism. In complex (PDB ID: 1XBB) of imatinib (Gleevec) and Syk (spleen tyrosine kinase), two regions are located in the binding sites including residues which form hydrogen bonds and van der Waals interactions with imatinib. Another one contains two phosphorylation sites, S579 and T582, although the role of those residues is not obvious.

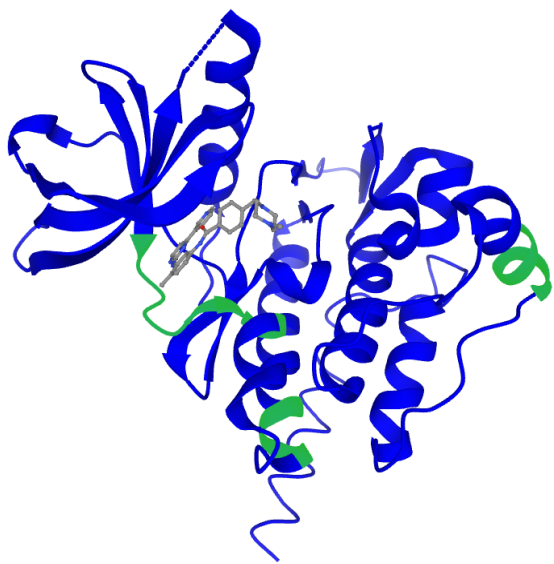

Fig 4. Complex (PDB ID: 1XBB) of imatinib (Gleevec) and Syk (spleen tyrosine kinase)

# Conclusions

Protein classification is important for biological research. This paper focuses on three critical problems on protein classification: multi-label CAZyme classification, protein graph classification and compound-protein interactions prediction. This paper proposes three models for protein classification. Firstly, this paper proposes a Multi-label CAZyme classification model using CNN-LSTM with Attention mechanism. Secondly, this paper proposes a variational graph autoencoder based subspace learning model for protein graph classification. Thirdly, this paper proposes Graph Isomorphism Networks and Attention-based CNN-LSTM for compound-protein interactions prediction. The proposed models are effective for protein classification. Ablation study is proposed to validate that graph isomorphism networks is superior to graph convolution and graph attention while incorporating Attention-based CNN-LSTM for compound-protein interactions prediction.

Source code and data are available at https://github.com/zshicode/GNN-AttCL-protein. Besides, this GitHub repository collects and collates the benchmark datasets for three problems mentioned above, including CAZyme classification, enzyme protein graph classification, compound-protein interactions prediction, drug-target affinities prediction and drug-drug interactions prediction. Hence, the usage for evaluation by benchmark datasets can be more conveniently.

Biological sequence and graph classification tasks are widely applied in bioinformatics research. The proposed universal methods for protein sequence and graph modeling, are with strong potentiality for other biological sequence and molecular graph classification tasks.